\documentclass[12pt,manuscript,flushrt]{aastex}
\usepackage{natbib}

\newcommand{\minusthree}{$^{-3}$}
\newcommand{\minusfive}{$^{-5}$}

\newcommand{\plusfive}{$^{5}$}
\newcommand{\OIII}{[\textsc{Oiii}]}

\newcommand{\kms}{\ km s$^{-1}$}
\newcommand{\galaxy}{NGC~4151}
\newcommand{\gal}{NGC~1068}
\newcommand{\HST}{\emph{HST}}
\newcommand{\fig}{Figure~}
\newcommand{\figs}{Figures~}
\newcommand{\point}{$ \!\!.  $\thinspace}
\newcommand{\vracc}{$ \mathrm{v=kr}  $}
\newcommand{\vrdec}{$ \mathrm{v=v_{  max  }-k^{  '  }r}  $}
\newcommand{\vrootracc}{$ \mathrm{v=k\sqrt r}  $}
\newcommand{\vrootrdec}{$ \mathrm{v=v_{  max  }-k^{  '  }\sqrt r}  $}
\newcommand{\rlaw}{$ r\ $law}
\newcommand{\rootrlaw}{$ \sqrt r\ $law}
\newcommand{\resolvingpower}{$ \lambda/\Delta\lambda $}
\newcommand{\roughly}{$ \sim $}
\newcommand{\approximately}{$ \approx $}

\shortauthors{Das et al.}  \shorttitle{Kinematics of the NLR in
NGC~1068}

\begin{document}

\title{Kinematics of the Narrow-Line Region in the Seyfert
  2 Galaxy \gal: Dynamical Effects of the Radio Jet\altaffilmark{1}}

\author{V. Das\altaffilmark{2}, D.M. Crenshaw\altaffilmark{2},
  S.B. Kraemer\altaffilmark{3}, R. P. Deo\altaffilmark{2}}

\altaffiltext{1}{Based on observations made with the NASA/ESA Hubble
  Space Telescope. STScI is operated by the Association of Universities
  for Research in Astronomy, Inc., under NASA contract NAS5-26555.}

\altaffiltext{2}{Department of Physics and Astronomy, Georgia State
  University, Astronomy Offices, One Park Place South SE, Suite 700,
  Atlanta, GA 30303, das@chara.gsu.edu, crenshaw@chara.gsu.edu, deo@chara.gsu.edu}

\altaffiltext{3}{Catholic University of America and Laboratory for
  Astronomy and Solar Physics, NASA's Goddard Space Flight Center, Code
  681, Greenbelt, MD 20771, stiskraemer@yancey.gsfc.nasa.gov.}

\begin{abstract}
  We present a study of high-resolution long-slit spectra of the
  narrow-line region (NLR) in NGC 1068 obtained with the Space
  Telescope Imaging Spectrograph (STIS) aboard \emph{The Hubble Space
    Telescope (HST)}. The spectra were retrieved from the Multimission
  Archive at Space Telescope (MAST) obtained from two visits and seven
  orbits of \HST\ time.  We also obtained MERLIN radio maps of the
  center of \gal\ to examine the dependence of the NLR cloud
  velocities on the radio structure. The radial velocities and
  velocity dispersions of the bright NLR clouds appear to be
  unaffected by the radio knots, indicating that the radio jet is not
  the principal driving force on the outflowing NLR clouds. However,
  the velocities of the fainter NLR clouds are split near knots in the
  jet, indicating a possible interaction. Biconical outflow models were
  generated to match the data and for comparison to previous models
  done with lower dispersion observations. The general trend is an
  increase in radial velocity roughly proportional to distance from
  the nucleus followed by a linear decrease after roughly 100 parsec
  similar to that seen in other Seyfert galaxies, indicating common
  acceleration/deceleration mechanisms.
\end{abstract}

\keywords{galaxies: kinematics and dynamics\,--\,galaxies:
  individual\\ (\gal) galaxies: Seyfert\,--\,AGN: emission
  lines\,--\,ultraviolet: galaxies} ~~~~~

\section{Introduction}
\gal\ is classified as a Seyfert type 2 active galaxy because of its
narrow (FWHM \approximately\ 500 \kms) permitted and forbidden optical
emission lines and its weak underlying non-thermal continuum. However
from spectropolarimetry studies, \citet{Antonucci} discovered a
featureless continuum and broad (FWHM \approximately\ 3000 \kms)
permitted lines in the core of \gal, similar to the spectrum of
Seyfert 1s. Their conclusion was that the central continuum source and
broad line region (BLR) in \gal\ is hidden behind a geometrically and
optically thick dusty torus. Subsequent studies by \citet{Miller}
found that four out of eight previously classified Seyfert 2 galaxies
reveled hidden Seyfert 1 nuclei and \citet{Kay1994} identified
additional Seyfert 1 nuclei in a spectropolarimetric study of a large
sample of Seyfert 2 galaxies. The unification model seems stronger
than ever after several more authors found hidden broad-line regions
in Seyfert 2 cores
\citep{Tran1992,Young1993,Young1996,Moran,Tran2001,Lumsden}.  More
recent spectropolarimetric studies by \citet{Zak} on Type II Quasars,
the luminous analogs of Seyfert 2 galaxies, further compliment the
unification model of active galactic nuclei (AGN). They found that
five out of the 12 studied objects showed broad lines.  Although it is
generally accepted that the two classes of Seyfert galaxies harbor the
same type of nucleus that are seen differently because of viewing
angle, it remains to be shown that they are indeed one and the same.
The ability of our models to match the data of both Seyfert types
provides yet another piece of compelling evidence for the unification
scheme.

The ionization mechanism of the NLR of Seyfert galaxies has been
scrutinized intensely over the past few decades. The two leading
mechanisms were photoionization \citep[see][and references
therein]{Koski,Ferland} and shock ionization \citep[see][and
references therein]{Dopita}. Photoionization was suggested as the
dominant source of energy to produce the emission lines seen in the
NLR as early as 1964 \citep{GreenSchmidt,Oke}, where it was shown that
shocks would not suffice to produce continuous energy in a small
region over large timescales. Models by \citet{Osterbrock} showed that
photoionization can account for both high and low ionization emission
lines seen in QSOs.

Recently, studies have shown that most of the emission lines seen in
the NLR can be accounted for by assuming a central ionization source
illuminating a multi-component gas \citep[and references
therein]{Kraemer2000a,Kraemer2000b,Groves}, although an additional
localized source of ionization in \gal\ was detected, which may be due
to shocks, at the point where the NLR gas decelerates. This kind of
shock may be caused by the interaction of NLR clouds with diffuse
plasma in the ambient medium, providing the drag in our kinematic
models \citep{Kraemer2000c}. \citet{Mundell} showed that the
misalignment of the radio jet and NLR in \galaxy\ supports
photoionization of the NLR by the AGN as the dominant excitation
mechanism, but that enhanced optical emission observed in a few clouds
bounding the radio jet might be shock related. Since it is now
generally accepted that photoionization is the dominant source of
ionization of the NLR gas, with shocks possibly contributing in
localized regions, we will focus our paper on kinematic models to gain
more insight into how these sources of energy may affect the
acceleration of the gas.

With the launch of \HST\ and its high angular resolution
(0\arcsec\point1), the NLR of Seyfert galaxies has received
considerable attention. With the long-slit capability of the Space
Telescope Imaging Spectrograph, and the Faint Object Camera (FOC),
detailed constraints on kinematic models of the NLR in the two types
of Seyfert galaxies became possible. In turn, these models provide
good diagnostics upon which dynamical analyzes can be based. One such
kinematic study of the NLR clouds is provided in this paper as a tool
for this endeavor. We used a systemic redshift of \gal\ of 1148 \kms\
from {\sc HI} measurements by \citet{Brinks} and a distance of 14.4
Mpc \citep{Bland-Hawthorne}, so that 0\arcsec\point1 corresponds to
7.2 pc.

\subsection{Lateral Expansion away from the Radio Jets}
Based on FOC long-slit spectra, several authors suggest that the NLR
clouds are accelerated by shocks generated by interactions between a
radio jet and the ISM, such that clouds are carried away laterally
from the jet axis by an expanding cocoon of shocked gas \citep[and
references therein]{Axon,Capetti}. In \citet{Capetti}, long slit FOC
spectra were taken perpendicular to the radio jet of Mrk 3. They found
that in some slits, the \OIII\ emission line velocities split into two
systems separated by several hundreds \kms\ apart on either side of
the jet. One of the systems is receding and the other one is
approaching relative to the host systemic velocity, which
\citet{Capetti} interpreted as expansion of gas away from the jet. In
\citet{Axon} the same splitting of velocities across a radio jet was
observed in \gal\ and also interpreted as expansion of the gas away
from the radio jet. The jet-driven models suggest that the lateral
expansion of the shocked cocoon will produce both blueshifted and
redshifted velocities with approximately equal magnitudes along the
jet/bicone axis at each location in the NLR, regardless of axis
orientation. A clear representation of this picture is shown in \fig6
in \citet{Nelson}.

The above model fails to explain the velocity structure in the NLR of
\galaxy, where STIS spectra revealed that primarily blueshifts were
observed on the southwest side of the bicone and primarily redshifts
on the southeast side \citep[and references therein]{Das}. We noted in
\citet{Das} that acceleration of some co-spatial clouds started well
before any radio knot, and extended well across the knot, making it
very unlikely that the knots accelerate these clouds. Furthermore, in
\citet{Das}, we found many more \OIII\ knots than radio knots in
\galaxy\, and the clouds seem to accelerate even in the absence of any
radio material, and in cases where there is radio material, the
velocities of the clouds seem to be unaffected. We will see in this
paper that there are possible interactions between radio and \OIII\
knots in \gal, but that these interactions are modest.

\subsection{Radial-Outflow Models}
Evidence of radial outflow in the NLR of \gal\ was first presented in
ground based studies by \citet{Walker}, and subsequently by other
researchers \citep[e.g,][]{Arribas,Cecil}. Evidence for a biconical
structure is expected from a simple unified model, due to collimation
by a thick torus, \citep{Antonucci}, and was seen by \citet{Pogge}. A
sample of Seyfert 1s and 2s were compared to test the similarity of
both NLR morphologies in \citet{Schmitt}; they found biconical
structures in most of their Seyfert 2s. Both \citet{Schulz} and
\citet{Evans1993} have modeled the NLR of \galaxy\ and found it to
consistent with a biconical geometry.

Biconical outflow modeling of the NLR using STIS data was previously
done by our group for \gal\ (Seyfert 2), Mrk~3 (Seyfert 2), and
\galaxy\ (Seyfert 1). The NLR of \gal\ was modeled with data taken
with the G430L grating of STIS and the radial velocity measurements
were made using the bright \OIII\ $ \lambda $5007 emission with a slit
at position angle 202\arcdeg\ and 0\arcsec\point1 wide
\citep{Crenshaw1068}. Acceleration of the gas to \roughly\ 100 pc and
subsequent deceleration back to systemic velocity was clearly
seen. The model consisted of a simple biconical outflow geometry which
is evacuated along the axis and tilted slightly out of the plane of
the sky to match the amplitudes of blueshifted and redshifted
clouds. Blueshifted and redshifted clouds were observed in both the
northeast and southwest parts of the bicone, which explains the
``line-splitting'' observed by \citet{Axon}. The NLR of Mrk~3 was
modeled by \citet{Ruiz2001}, using a combination of STIS medium
resolution slitless and low resolution long slit data, and a similar
geometrical model matched the data well.  Clouds were seen to
accelerate close to the nucleus, then decelerate back to systemic
velocity, a trend which is seen in other Seyfert galaxies. In
\citet{Crenshaw4151}, the NLR of \galaxy\ was similarly modeled with
two slit positions, each 0\arcsec\point1 wide, at position angles
221\arcdeg\ and 70\arcdeg. The axis of the bicone was tilted to match
the blueshifted and redshifted clouds, which contrary to \gal,
occurred exclusively in the southwest and northeast bicones
respectively. With an inclination of the bicone axis with respect to
the plane of the sky of 40\arcdeg, the models fit the data reasonably
well. In \citet{Das}, we modeled the NLR of \galaxy\ using five
parallel slits at much higher spectral resolution (\resolvingpower\
\roughly\ 9000). The slits were each 52\arcsec$ \!  $ x
0\arcsec\point2 positioned at 58\arcdeg\ and covering most of the
NLR. The data allowed multiple \OIII\ components to be detected and
measured. The previous model from \citet{Crenshaw1068} was then
fine-tuned to allow a better fit. The models were also consistent with
the different slit position angles.

In an analysis of 10 Seyfert galaxies by \citet{Ruiz2005}, we studied
the kinematics of the NLR using slitless spectroscopy at high spectral
resolution. The sample includes eight Seyfert 2, one Seyfert 1.9, and
one Seyfert 1 galaxy. Each was observed with the STIS 1024 x 1024
pixel CCD detector through an open aperture (52\arcsec$ \!$ x
52\arcsec) and dispersed with the G430M grating.  A high resolving
power of \roughly\ 9000 allowed us velocity measurements down to
\roughly\ 33 \kms\ in the vicinity of the \OIII\ $\lambda$5007
line. Each target was observed during a single \HST\ orbit. Although
modeling was not done, the tell tale signs of
acceleration/deceleration of the gas seemed to be present in almost
all of their Seyferts. In all the above papers, the observations and
models indicate that the dominant accelerating mechanism emanates from
within tens of parsecs from the central engine and that the
acceleration is radial rather than lateral (from the jet axis).

\subsection{Kinematics of the NLR of \gal\ at High Spectral Resolution
\label{kinematics}}
Observations of \gal\ were taken on 1999 September, 1999 October, and
2000 September by \citet{Cecil}. They used STIS aboard \HST\ to obtain
spectra with the G430M grating, at a spectral resolution of
\resolvingpower\ \approximately\ 9000 and a spatial resolution of
0\arcsec\point1 along and 0\arcsec\point2 across each slit. Due to
complications with guide stars, the observing time was split into two
parts, one year apart, which resulted in a total of seven parallel
long slit spectra at position angle 38\arcdeg. Slits 1--5 were taken
in 1999, and slits 6--7 were taken the following year. However slits 5
and 6 contained the same region, except for an offset along the
slits. The spectra from each slit had a spectral coverage of 4820-5100
\AA, to obtain maps of the \OIII\ and H$ \beta $ profiles \citep[for
more details see][]{Cecil}. An additional observation with the G430M
grating (slit 8) was taken by Antonucci (PI) in 2000 January at
position angle of 10\arcdeg.

\citeauthor{Cecil} analyzed several complexes and knots of gas that
cover the entire NLR. These were sampled into radial velocity profiles
with some knots showing radial velocities as high as -3200
\kms. Simple biconical outflow models were overlaid on their \OIII\
spectral images for all slit positions, and velocities were
compared. They also overplotted model velocities where the gas is
allowed to expand away from the radio jet axis. They found that a
higher maximum velocity was required to better match the blue wing of
the northeast bicone. The red wing was not matched. The lateral
expansion model also predicted larger velocities for the red wings
than the data. They suggest that the redshifted bright emission seen
in the northeast cone is caused by the clouds being pushed into the
galactic disk by the expansion of the radio lobe, and that these
clouds are most likely dragged along the galactic disk. This may
explain the clouds's lower velocities. The blueshifted clouds are
accelerated away from the galaxy and suffer no such fate and hence
their higher velocities. They also suggest that the high-velocity
knots in the northeast can be explained by ablation of massive
infalling dusty clouds.

We have remeasured \citeauthor{Cecil}'s data specifically for testing
simple velocity models, by separating the multiple emission components
seen along the slit. As in \citet{Das}, we wanted to test the
relations among the different components and also to check for unusual
flow patterns of the faint clouds as seen in the NLR of \galaxy. We
also wanted to test for radio/\OIII\ interactions by comparing
co-spatial slit coverage of optical emission data from \citet{Cecil}
and radio emission from \citet{Gallimore2004}.

\section{Observations\label{Observations}}
The top panel of \fig\ref{intensity} shows the slit placement of
\citet{Cecil} on top of an FOC \OIII\ image of \gal\ from
\citet{Macchetto}. The slits cover an area of 86.4 parsec across and a
few thousand parsecs along the slits. However, the brightest emission
is seen within a few hundred parsecs of the central engine
(CE). Roughly 90\% of this NLR emission is covered by the slits with
some emission seen outside of slits 1 and 7. We also include slit 8,
which is made up of three archival G430M-grating exposures taken in
2000 January and first analyzed by \citet{Cecil} (see
\S\ref{kinematics}). We also overlay a high-resolution MERLIN radio
map taken on 1998 January 2. The map was provided in its fully reduced
form by \citet{Gallimore2004}. With accurate astrometry on the radio
data, we can test the correlation between the radio knots and the
\OIII\ clouds, as was done in \citet{Das}. We previously found no
significant radio interference on the NLR clouds in \galaxy, such as
expansion away from the jet axis, as would be indicated by redshifts
and blueshifts of similar magnitude at the radio knot positions.

The high resolution radio map from \citet{Gallimore2004} was scaled to
match the pixel resolution of the \OIII\ image from \citet{Macchetto}.
We then followed the \HST\ observation procedure for aligning the
slits. At the time of observation, slit 4 was centered on the
continuum hot spot, and the rest of the slits were offset in the
dispersion direction by 0\arcsec\point2 (the slit width). Slit 8
(taken by Antonucci) was oriented at a different position angle but
was also centered on the hot spot. The intersection of slits 4 and 8
mark the position of the hot spot in \fig\ref{intensity} (this figure
is in \OIII\ emission and hence the continuum hot spot is not
visible). We then aligned the radio and the \OIII\ images by placing
radio knot S1 0\arcsec\point13 $\pm$ 0\arcsec\point1 South and
0\arcsec\point02 $\pm$ 0\arcsec\point1 West of the continuum hot spot.
Our motivation stemmed from absolute astrometry studies by
\citet{Capetti1997} who placed the optical continuum peak at $\alpha$
= 02$^h$42$^m$40.711$^s$, $\delta$ =
-00\arcdeg00\arcmin47\arcsec\point81 (J2000) with a precision of 80
mas, and by \citet{Muxlow} who placed the radio component S1 at
$\alpha$ = 02$^h$42$^m$40.7098$^s$, $\delta$ =
-00\arcdeg00\arcmin47\arcsec\point938 (J2000) with precision of 20
mas. The central engine, which is presumably the location of the
supermassive black hole (SMBH), is now generally accepted to be
positioned at S1 because of the observed thermal inverted spectrum
between 1.3 cm and 6 cm, and also because of the H$_2$O maser emission
centered on that location
\citep{Gallimore1996,Greenhill1997,Kumar1999,Beckert2004}. A more detailed
treatment of the locations of these and other features in various wave
bands is presented by \cite{Galliano}.

The bottom panel of \fig\ref{intensity} shows an enlargement of the
top panel. The UV and optical continuum hot spot and also NLR-cloud B
are found to be coincident within 25 mas; therefore the location of
the continuum (`hot-spot'), shown by the white arrow, is near the dark
spot that is NLR-cloud B in this image, in the centers of slits 4 and
8. The radio axis after the bend in the radio jet, is nearly aligned
with our slits, extending from knots C to S1. This allows for
one-to-one comparison between the radio knots and the \OIII\ clouds.
Slits 3, 4, and 5 cover most of the radio region and slit 8 captures
radio knots S1, S2, and C almost perfectly.  It is then easy to
extract similar regions of \OIII\ and radio data using the slits as a
guide.

\fig\ref{slit4image} shows a fully reduced spectral image of the
\OIII\ $\lambda\lambda$4959, 5007 emission as it appears through slit
4. The line across the image pointed at by the arrow is the ``hot
spot" which is presumably produced by nuclear light scattering off of
electrons \citep{Crenshaw1068a}. The hidden nucleus is then
0\arcsec\point13 south of this optical hot spot and is placed at
0\arcsec\ in the image, (see \S\ref{Observations}).

\section{Analysis\label{Analysis}}
Along each slit we extracted one spectrum per pixel in the
cross-dispersion direction (\roughly0\arcsec\point05/pix) and fitted
the identifiable component peaks associated with the \OIII\
$\lambda$5007 line with a local continuum and multiple Gaussians. We
measured the radial velocities for each component based on the centers
of the Gaussian fits. Because we are resolving the \OIII\ knots along
a slit, we often obtain multiple measurements of the same
component. Also the radial velocity may change gradually across a knot
of \OIII\ gas, suggesting that the knots themselves are made up of
smaller clumps or that there are velocity gradients across them. A
section of slit 4 shown by the dashed lines across
\fig\ref{slit4image} is extracted and fitted with Gaussians as an
example. This progression of spectra is shown in
\fig\ref{spectra}. The different components for each spectrum are
color coded according to the total flux in the line at each location:
red for highest flux, blue for medium flux, and black for lowest
flux. Such distinction of the components was necessary to compare our
models to previous ones in \citet{Crenshaw1068}, since mostly the
bright and medium flux knots were detected in their data. Also in
\citet{Das}, the faintest component was found to behave differently
than the overall trend of the bright and medium flux components in
\galaxy. We tested for similar flow patterns of the clouds in this
paper by tracing the different components. We also measured the
position in arcseconds of all components along a slit relative to the
center of that slit\footnote{The center of any slit is defined as the
position offset in the dispersion direction from the equivalent value
of the pixel that corresponds to the position of the CE in slit 4.}.

\figs\ref{vel_fwhm_flux1}, \ref{vel_fwhm_flux2}, and
\ref{vel_fwhm_flux3} show the kinematic components in plots of radial
velocities, FWHM, and fluxes as a function of projected position from
the center of each respective slit. The colors represent the different
components as explained previously. In all the radial velocity plots,
data points near zero velocity in the range 2--10 arcseconds are in
the plane of the galaxy as indicated also by their low FWHM. These
points do not play any role in our modeling and are shown only for
completeness. For points with larger FWHM, the general trend is an
increase in radial velocity from the origin to $\sim$ 1\arcsec\point5,
followed by a decrease to the systemic velocity at $\sim$ 4\arcsec.
The fluxes generally decrease away from the CE, with two spikes, one
at \roughly3\arcsec\point5 and one at \roughly8\arcsec, indicating the
presence of more gas at these locations, which can be seen in the
\OIII\ images of \citet{Evans}. The FWHM does not present any
discernible trends in any of the colors except for a slight decrease
away from the CE.

For the radio comparison, we first scaled the radio to match the
resolution of the \OIII\ map. The radio contours were then overlaid on
the \OIII\ map (\fig\ref{intensity}) to locate the slits that
intersect the radio region. The boundaries of the intersecting slits
were then used to extract regions along slits 3, 4, 5, and 8 from the
radio map. These slits were summed across their widths to find the
total intensities along each slit. The radio intensities and the
\OIII\ data were then plotted against position for each slit, and then
compared for any \OIII\ disturbance. These results are presented in
\S\ref{Results}.

Estimates of errors in our measured radial velocities come from three
sources. 1) By realizing that the components are not perfect
Gaussians, we measured the actual centroid for the $ \lambda $5007
lines of a random sample of spectra and found a maximum fitting error
of 0.89 \AA. 2) The finite slit width of 0\arcsec\point2 introduces an
uncertainty of 0.56 \AA\ for all clouds that are dispersed. 3)
Finally, we are limited to how far away from the hot spot we can
measure fluxes because of the rapidly degrading signal-to-noise (S/N)
ratio. We measured random noisy spectra several times and find that
the wavelength shifts vary by \roughly0.25 \AA. All these are added in
quadrature and converted into velocity to give a maximum error of
$\pm$65\kms. In all fits, the continuum placement is chosen by eye. We
varied the placement of the continuum for randomly selected spectra,
but find that the fit is not significantly affected. The S/N ratio of
the \OIII\ $\lambda$5007 line rapidly decreases away from the hot spot
and we did not measure the emission if the S/N was $ \lesssim $ 3,
which typically occurred at distances of $ \gtrsim $ 6\arcsec. There
are also some errors explicit to the separation criteria due to
occasional mixing between the medium and low flux components; however
such misidentification of the components is small, (see \citealt{Das}
for a full explanation).

\section{Kinematic Models\label{Models}}
In \citet{Crenshaw1068}, \citet{Crenshaw4151}, and \citet{Ruiz2001},
models were developed to match the NLRs of \gal, \galaxy, and
Mrk~3. The low-resolution STIS data provided measurements of only the
bright NLR clouds and limited spatial coverage of the NLR, since only
one or two slit positions were obtained. Therefore, in \citet{Das}, we
reexamined the previous models for \galaxy\ with high-resolution STIS
data and full coverage of the NLR. Again in this paper, we use higher
dispersion STIS data with large coverage of the NLR to map the
kinematics of \gal\ and test previous biconical outflow models. Model
parameters that we use are listed in Table \ref{model-parameters}. The
parameter z$ _{ max } $ is measured along the bicone axis from the
apex of the model to one end of the bicone (the bicone is symmetrical
in geometry about the apex). The parameter v$ _{ max } $ is the
maximum velocity attained by the outflow with respect to the nucleus,
and the parameter r$ _{ t } $ is the distance from the nucleus to
where the velocity of outflow reaches its maximum. The inner and outer
opening angles are actually half-angles, as they are measured from the
bicone axis to the inner and outer edges of the bicone
respectively. The inclination of the bicone axis is measured with
respect to the line-of-sight (LOS) and is zero if the bicone axis is
perpendicular to the LOS. Since the \OIII\ outflow velocity seems to
be following some function-dependent relationship with distance from
the nucleus, we used simple velocity laws for the acceleration and
deceleration phase of the gas; the ones that worked best were \vracc\
and \vrdec\ respectively, which we collectively refer to as the
\rlaw. We tested other velocity laws of the form $ \mathrm{v\propto
r^{ n }} $ but most were not acceptable. One other law matched the
data well, our so called \rootrlaw, where we used \vrootracc\ for the
acceleration and \vrootrdec\ for the deceleration phase of the
gas. The distinction between the two velocity laws is subtle and can
be seen in \fig9 in \citet{Das}. We continue to use the \rlaw, also
referred to as the `Hubble Flow' law by many, in this paper.

Our previous procedure, described in more detail in our past papers,
first generated a 2-D radial velocity map based on parameters given in
Table \ref{model-parameters}. For a given parameter set, the velocity
field was generated for every increment of angle between the inner and
outer opening angles. Normally we used an increment of
1\arcdeg. Examples of such fields are shown in \figs5 and 6 in
\citet{Crenshaw4151}. Samples of radial velocities were then extracted
from the model along slit positions, orientations, and widths that
matched those of the observations. These velocities were then plotted
against position from the center of the respective slit, together with
the data points. The parameters were then adjusted slightly and the
entire procedure was repeated until we matched the plots of data and
model as best as possible, whereby the final parameters are quoted as
our best model fit.

In previous models, we used the continuum hot spot as the apex for the
biconical outflow of the gas. In this paper we modified the starting
point of the outflow to correspond to the position of the hidden
nucleus located 0\arcsec\point13 south of the hot spot
\citep{Galliano}. Although such changes do not affect our final model
parameters significantly, we decided to use this location because it
is most likely the site of the CE and it might be useful
for future dynamical models. Also we want to align as best as possible
the radio knots with the positions of the \OIII\ clouds, relative to
the true center. Such alignment is crucial in testing the radio
jet/NLR interactions and the acceleration mechanism of the gas.

\label{test}In this paper we decided to use a 3-D modeling scheme
which is more efficient and greatly reduces our run-time compared to
our previous models. In addition the reader can more easily picture
and comprehend the biconical outflow geometry. We start by filling in
a cubic array with velocities for a given parameter set and velocity
law. Note that we fill the cubic array with radial velocities as would
be measured by the observer. Such velocities are first determined by
our velocity law of outflow and are then converted to radial
velocities depending on their positions or coordinates in the model
cube. For the purposes of illustration, the radial velocities are then
assigned a color based on the amount of redshift or blueshift.
Examples of such models are shown in \fig\ref{models}. We then take
slices through the cube corresponding to slit positions, orientations,
and widths that match those of observations. Velocities and positions
are then extracted along the slices and plotted together with
positions and radial velocities from the observations. Panel (a) in
\fig\ref{models} show the velocity field at an inclination of
0\arcdeg. The parameters that were used to generate the bicone are not
the best fit values. They were chosen just for image clarity purposes.
Panels (b) and (c) give a clear view of the front and back surfaces
respectively of bicone (a).  The colors red and blue represent the
redshift and blueshift of the surfaces and the key is displayed in
panel (g). The darkest blue and red colors in these panels represent
the turnover point. This is where the \OIII\ gas is at its maximum
velocity after which it decelerates back to systemic velocity. Due to
the upright orientation of the bicone, gas outflows from its apex are
symmetrical above and below.  The expected symmetry of the velocities
is represented by the equal color gradients on both front and back
surfaces of the bicone. The situation changes when we apply an
inclination to the bicone model as in panels (d) and (e). Now the
inclination is 5\arcdeg, with the top toward the observer. Panel (d)
is shown with a shaded cross-cut plane that represents the position,
orientation, and width of slit 4. One can see slightly higher
blueshifts in the top right and slightly higher redshift in the bottom
left of (e). The cross section of this plane is shown in panel (e). At
each point going up along (e), we extract the range of velocities
across the figure and plot those velocities correspondingly in panel
(f). Panel (f) represents our final stage for comparing model with
real data points. At this stage the model slit as represented by (f)
and data points from the corresponding slit are compared. If the fit
is found to be unsuitable, the parameters are adjusted slightly and
the entire process of generation (d), slit extraction (e), and
velocity extraction and plotting (f), undergoes repeated runs until we
determine the best match to the data.

\section{Results\label{Results}}
\figs\ref{plotshade1} and \ref{plotshade2} show the best fit
comparison plots for all eight slits. The models are in shaded gray
and the data points in colors, and are plotted with positions relative
to the center of the respective slits. These models were done with our
best fit parameters listed in Table~\ref{compare-ngc1068}.  Previous
model parameters found in \citet{Crenshaw1068} are also listed in that
table for comparison.  Our results are strikingly consistent with our
earlier ones, except that we increase the maximum extent of the bicone
and the maximum velocity at turnover point, consistent with the
results of \citet{Cecil}. In addition, the position angle of the
bicone was changed from the previously accepted value of 15\arcdeg\
\citep{Evans} to 30\arcdeg, consistent with values derived by
\citet{Bergeron} and \citet{Pogge}. This change gave a coherent fit
across all eight slits. Slits 5 and 6 cover the same region except for
an offset along the slit. They were taken one year apart and show
almost identical data points. The difference is seen mostly in the low
and medium flux clouds, which can be attributed to our separation
criteria for distinguishing the different kinematic components (see
\S\ref{Analysis}).

The fit to the bright and most medium flux clouds is relatively good.
Note that the long line of red points near zero velocity in the
northeast are points in the host galaxy which were not fitted. These
points are actually low absolute flux and low FWHM points detected
with only one kinematic component per location. Because of this, our
separation routine assigned the color red to these points, but clearly
these points are not in any outflow pattern. There are also some faint
clouds outside of the bicone geometry around zero position in all
slits that are not fit. We present an explanation in terms of the
radio jet for this sort of flow pattern of the fainter clouds in
\S\ref{Discussion}.

The trend in all slits is that the points accelerate then decelerate
with approximately linear velocity laws. The sparsity of points
enclosed by, but not in, the shaded regions of \figs\ref{plotshade1}
and \ref{plotshade2} indicate that the bicone interior is hollow.
Slit 8 does not show this hollowed out area as the rest of the slits,
because it is placed at a position angle of 10\arcdeg\ (at opposed to
38\arcdeg\ for the rest of the slits), which makes it lie along one
edge of the bicone. For all the slits, the data follow the models
quite well in the northeast, even for slit 8, where the data seem to
agree with the filled shaded configuration reasonably well.

The lack of data points in the southwest can be explained by the
shielding of the bicone by the host galaxy, as shown in
\fig\ref{host_galaxy}. Our model predicts that the outer half-opening-
angle of the bicone is 40\arcdeg, with the northeast side inclined at
5\arcdeg\ out of the plane of the sky. This decreases the opening
angle of the southwest side relative to the plane of the sky by 5
degrees. The host galaxy is inclined by 40\arcdeg\
\citep{deVaucouleurs}, thus effectively placing more material between
us and the southwest side of the bicone. The northeast side of the
bicone is therefore above the galactic disk. Studies of
\textsc{HI} absorption in NGC~1068 \citep{Gallimore1994} find
that the NE radio jet is also in front of the galactic disk, which is
consistent with our result. Both the northeast far side, and the
southwest near side of the bicone are flowing in/near the galactic
plane. Such interaction between gas in the host galaxy and ionizing
radiation in the bicone is responsible for carving out the morphology
of the Extended Narrow Line Region (ENLR).

\fig\ref{radio_velocity} is a plot of relative radio intensities along
slits 2, 3, 4, and 8, together with \OIII\ radial velocities within
2\arcsec\ of the CE. In panel 1 (slit 3) of this figure, the radio
knot (S2) does not seem to have any effect on the bright cloud
velocities, i.e., the bright points are not convincingly split into
redshifts and blueshifts at S2. Furthermore the narrow range of the
radio emission cannot account for the slowly but steadily increasing
blueshifted velocity going from 0\arcsec\point5 to
-0\arcsec\point5. At radio knot S1 in panel 2 (slit 4), there seem to
be splitting of the velocities, but this splitting extends all the way
beyond -0\arcsec\point2, where there is no radio. The splitting of the
bright clouds continues after radio knot NE, going to the
right. Similar scenarios apply to panels 3 and 4; the bright clouds
split in velocities, but not preferentially at the position of the
radio knots.

The fainter clouds at the positions of the radio knots in
\fig\ref{radio_velocity}, however, portray a different picture. The
green circles represent pairwise velocity splitting where the radio
knots are strongest. The split velocities have approximately equal
magnitudes in redshifts and blueshifts, and the splitting is seen only
in the low and medium flux clouds, but not in the high flux clouds. At
radio knot C in panels 2, 3, and 4, there is minimal splitting, with
low velocities. This could be due to the fact that the radio jet bends
at this position, after colliding with \OIII\ cloud C. The blue
circles represent faint clouds for which there is no corresponding
blueshifted velocity split. These pairwise splitting of velocities is
explained in terms of lateral expansion of the radio jet in
\S\ref{Discussion}.

\section{Discussion\label{Discussion}}
We compared our previous kinematic models for NGC 1068 with our
current ones in this paper and found that they remained consistent.
Radial outflow of the \OIII\ clouds can explain the general
large-scale flow pattern of the narrow-line regions of both types of
Seyfert galaxies, such as \galaxy\ from \citet{Das} and \gal\ from the
current paper. However the models are not a perfect match, and further
insight can be gained from the discrepancies seen in
\figs\ref{plotshade1} and \ref{plotshade2}. Superimposed on the
general trend of the flow, there are a number of clouds with peculiar
velocities not in the shaded regions of our model. Points near
systemic velocity that extends from \roughly2\arcsec\ to
\roughly9\arcsec\ are faint clouds in the host galaxy, on the far side
of the northeast cone, as shown in \fig\ref{host_galaxy}.

High-velocity points near 0\arcsec\ are also not in the shaded region
of our biconical model and may not be affected by processes that drive
the outflow of the bright clouds along the bicone. These points can
possibly be explained by lateral expansion of hot gas away from the
radio jet, as claimed by \citet{Axon} and \citet{Capetti1997}.
\fig\ref{jet_expansion} shows a schematic of this lateral expansion of
fainter clouds away from the jet/bicone axis. This flow however is
restricted to clouds within a few tens of parsecs from the CE, along
the region of the radio jet. The large scale outflow of the bright
clouds cannot be accounted for by the radio, and needs a different
acceleration process. The evidence for a small scale lateral flow was
previously shown in \fig\ref{radio_velocity} as velocity splitting of
the faint clouds at the positions of the radio knots, in
\S\ref{Results}. Another possible explanation is that these clouds may
be driven by radiation escaping through a patchy torus or by scattered
light, which would also explain their ionization outside of the
classic bicone structure \citep[e.g., see][]{Kraemer2000b}.
\citet{Schmitt2006} has shown that these clouds indeed have a lower
ionization state than those in the bicone, indicating a patchy torus
or at least a much lower extinction column than that in the LOS.
Recent models of the IR emission from the putative torus in Seyfert
galaxies also indicate a patchy geometry \citep[e.g.,][]{Nenkova2002}.

The large scale outflow itself is not well understood dynamically. The
gas seems to accelerate out to about 140 pc, then decelerate back to
systemic velocity. Such a flow pattern was observed in several other
Seyferts by our group. Mrk 3 shows a turnover in velocity at
\roughly80 pc and \galaxy\ shows turnover at \roughly96 pc. Whatever
is slowing the gas down is not known for certain, but we suspect that
it might be diffuse, possibly X-ray emitting, gas in the ambient
medium. The most likely driving mechanism, radiation pressure by line
driving, has some drawbacks. Models developed by \citet{Everett}
predict that the gas should reach high velocities very close in, at $
\lesssim$~1 pc. Gas is lifted off the accretion disk by
magnetocentrifugal and radiative forces, and then driven by
radiation. Part of the lifted gas is used for shielding, to avoid
complete ionization of the driven gas. The model, however, cannot
match our results because the observed acceleration of the gas is much
shallower than that predicted by the model, attaining maximum radial
velocity at \roughly100 pc, whereas the model predicts maximum
velocity within \roughly1 pc.

The large scale flow of the NLR clouds is not likely to be accelerated
radially by the radio jet. Considering that the density the radio
knots is of the order \roughly10\minusfive\ cm\minusthree and that of
the \OIII\ clouds is \roughly10\plusfive\ cm\minusthree, momentum
arguments and the large solid angle covered by the \OIII\ clouds would
make it difficult for the radio jet to push the clouds to high
velocity. \citet{Mundell} suggests that some NLR clouds are actually
responsible for bending the radio jet. For example, NLR knot C is most
likely responsible for the bend in the radio jet at radio knot C
\citep{Capetti}.

We have shown that the lateral expanding cocoon model
\citep{Axon,Capetti1997} cannot be the main driving mechanism for the
NLR clouds in NGC 1068. Furthermore, it cannot account for the
kinematics of the Seyfert 1 galaxies \galaxy\ \citep{Crenshaw1068} and
NGC 3516 \citep{Ruiz2005}. These galaxies unambiguously show redshifts
on one side and blueshifts on the other side the nucleus,
contradicting the lateral expansion model, which predicts redshifts
and blueshifts on both sides of the nucleus, regardless of inclination
of the bicones. However, the radio jet in itself could be responsible
for clearing a channel through the center of the bicone, providing the
hollowed out interior that matches our models, as stated in our
earlier papers. Also, the lateral expansion of the radio knots in
\gal\ may be responsible for pushing the fainter clouds aside, filling
in the apex in our biconical models with low ionization gas.

\acknowledgments We thank J.F. Gallimore for providing the high
resolution MERLIN radio map of \gal\ used in this study. Some of the
data presented in this paper were obtained from the Multimission
Archive at the Space Telescope Science Institute (MAST). STScI is
operated by the Association of Universities for Research in Astronomy,
Inc., under NASA contract NAS5-26555. This research has made use of
NASA's Astrophysics Data System.

%\clearpage
\bibliographystyle{apj}
\bibliography{apj-jour,paper}
\figcaption[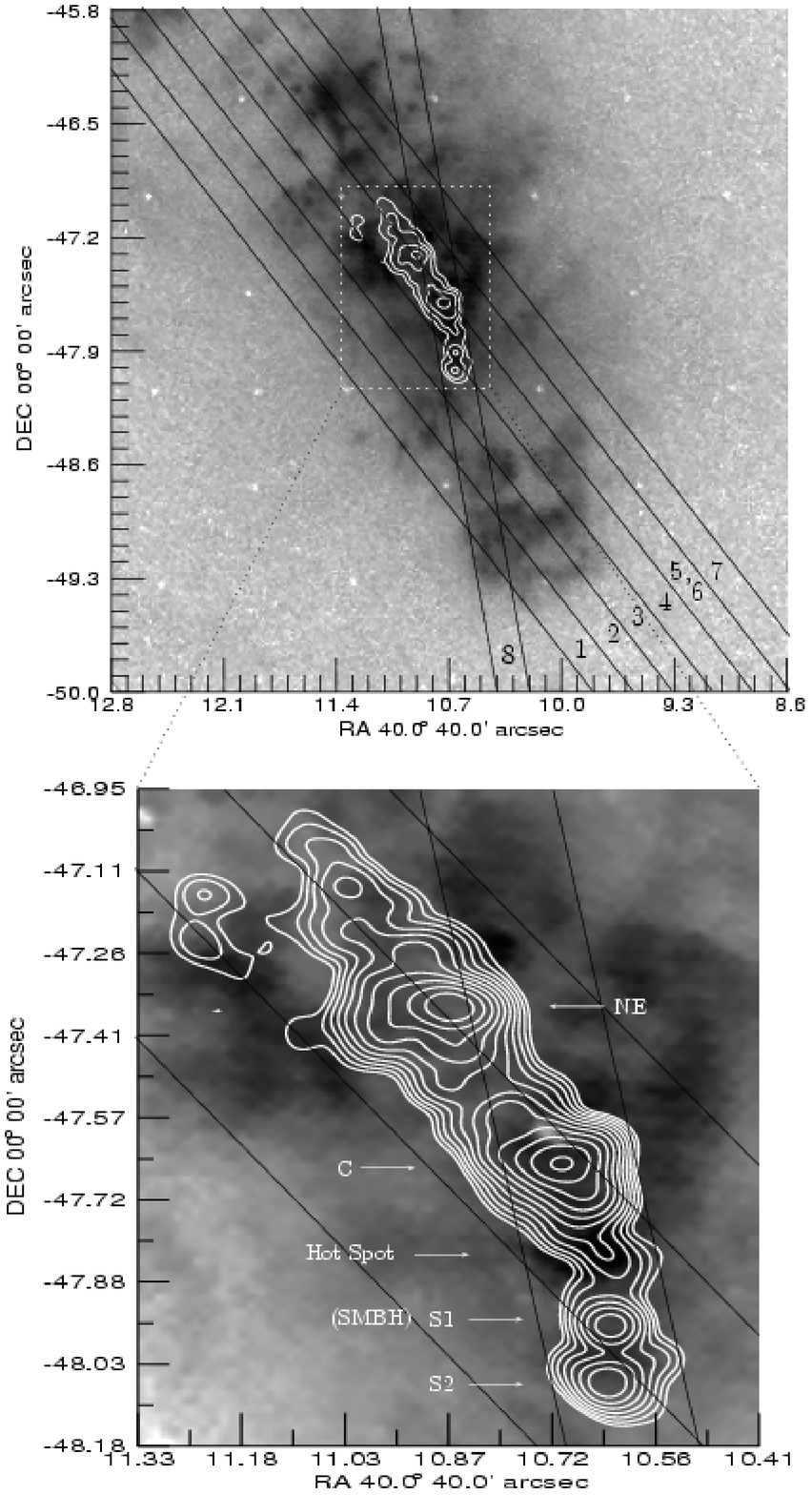]{Top: Faint Object Camera(FOC)/Corrective Optics
  Space Telescope Axial Replacement(COSTAR) log intensity \OIII\ image
  of the NLR of \gal\ shown with the orientation and position of the
  seven overlying slits. The image was taken through the narrow-band
  filter (F501N), and the slits are those of STIS observations. The
  position angle of slits 1--7 on the sky is 38\arcdeg, and slit 8 is
  at 10\arcdeg. Contours are from the radio image. Bottom: Enlarged
  view of the white box from top showing the radio region with the
  slits overlaid.  Note that slits 3, 4, and 5 cover the strong
  radio knots, while slit 8 redundantly covers radio knots C, S1, and
  S2. The center of the continuum hot spot is found at the
  intersection of slits 4 and 8 as shown by one of the white arrows,
  while the other white arrows point to known radio knots. Radio knot
  S1 marks the position of the SMBH.\label{intensity}}
\figcaption[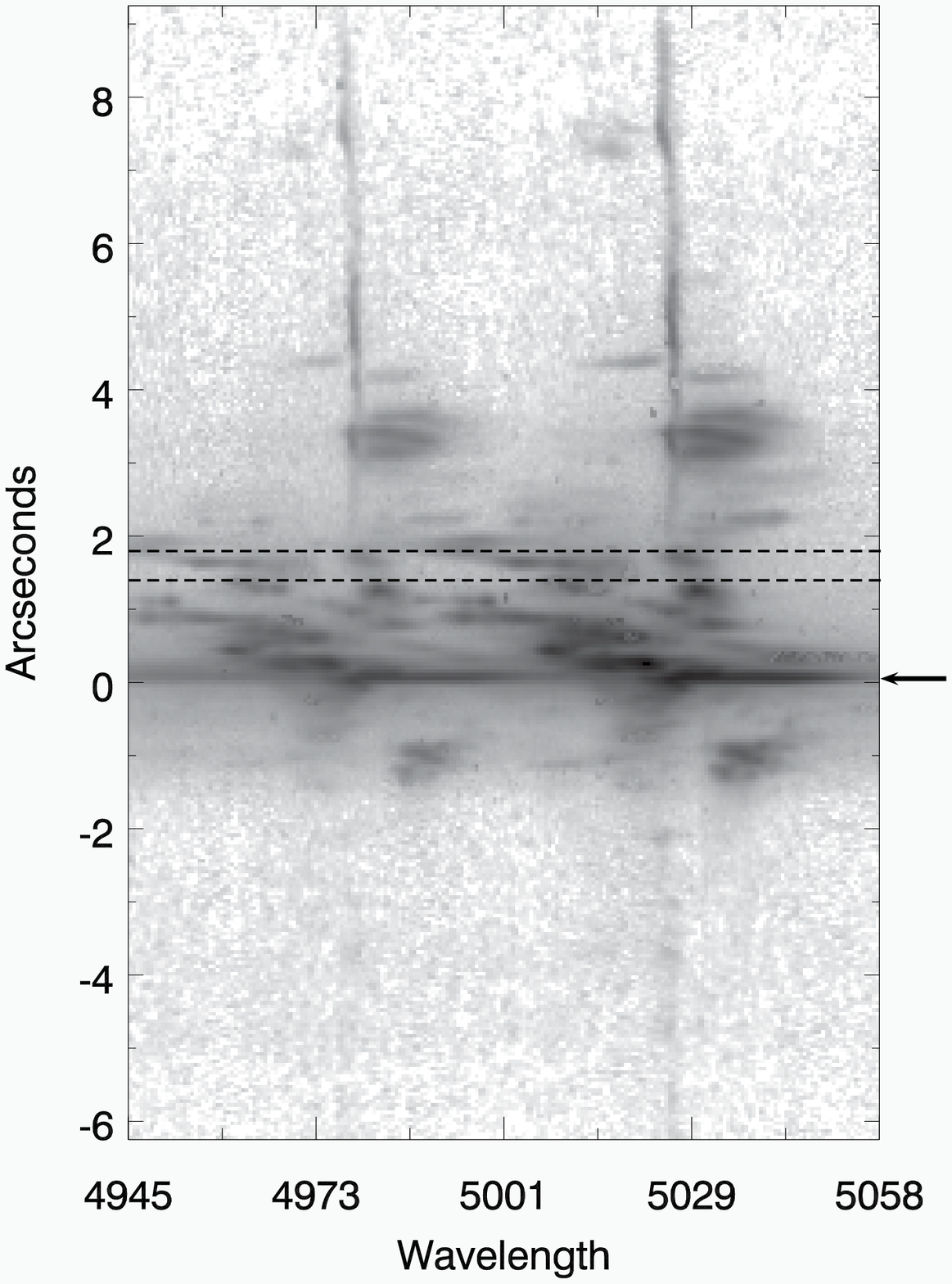]{\OIII\ $ \lambda $4959,5007 dispersed image of slit
  4. The arrow points to the optical continuum hot spot. The hidden nucleus is
  positioned at zero arcsecond slightly below the continuum. The dotted
  lines across the image represent the region of extraction for the
  progression of spectra in \fig\ref{spectra}. This progression goes up
  the slit. \label{slit4image}}
\figcaption[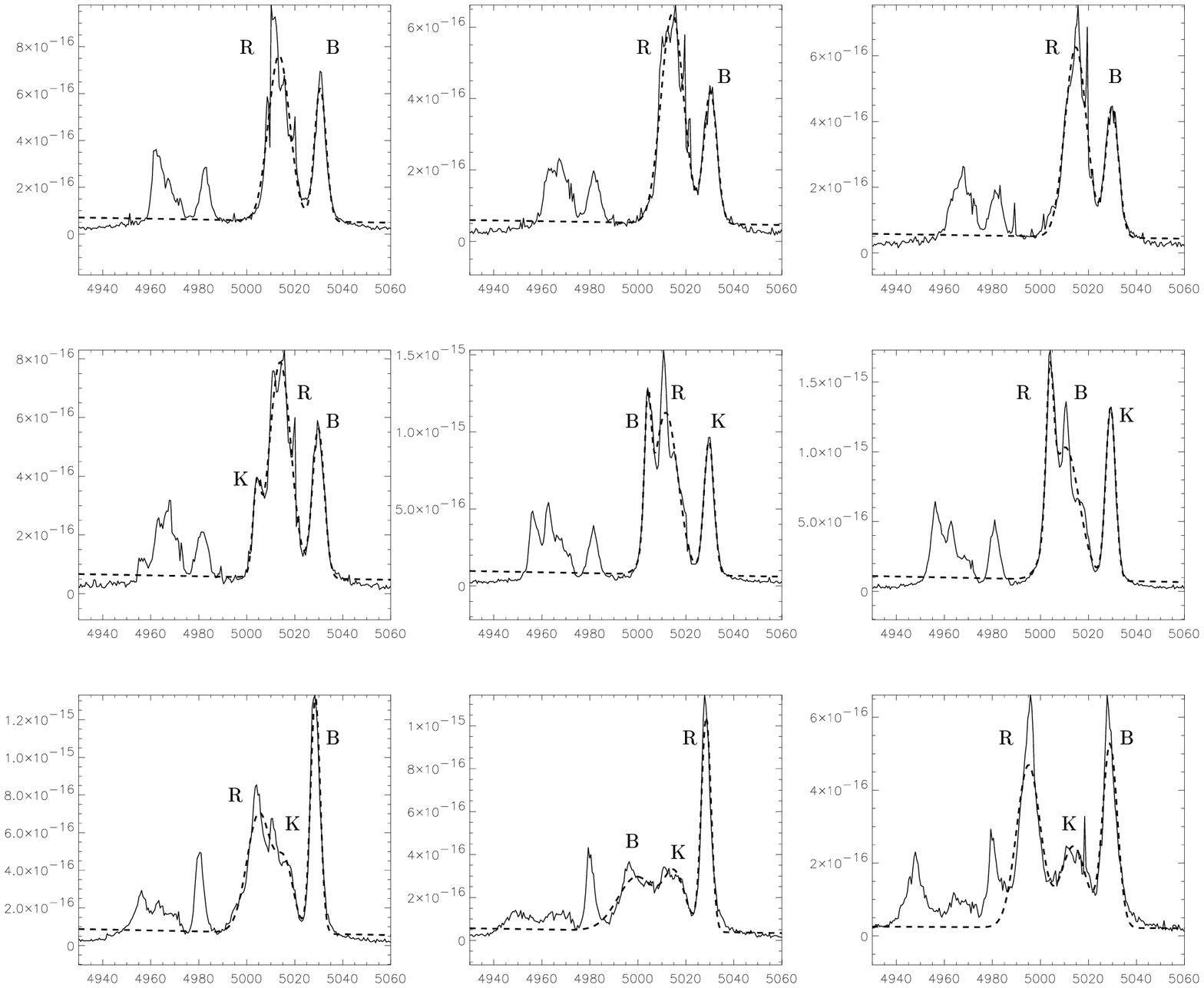]{Progression of spectra from slit 4 (see
  \fig\ref{slit4image}) showing the fluctuation of \OIII\ components
  with position along the slit, together with the summed gaussian
  fits. These components are separated
  according to total relative flux in the lines and are color coded
  red (R), blue (B), and black (K), for high, medium, and low flux,
  respectively. \label{spectra} }
\figcaption[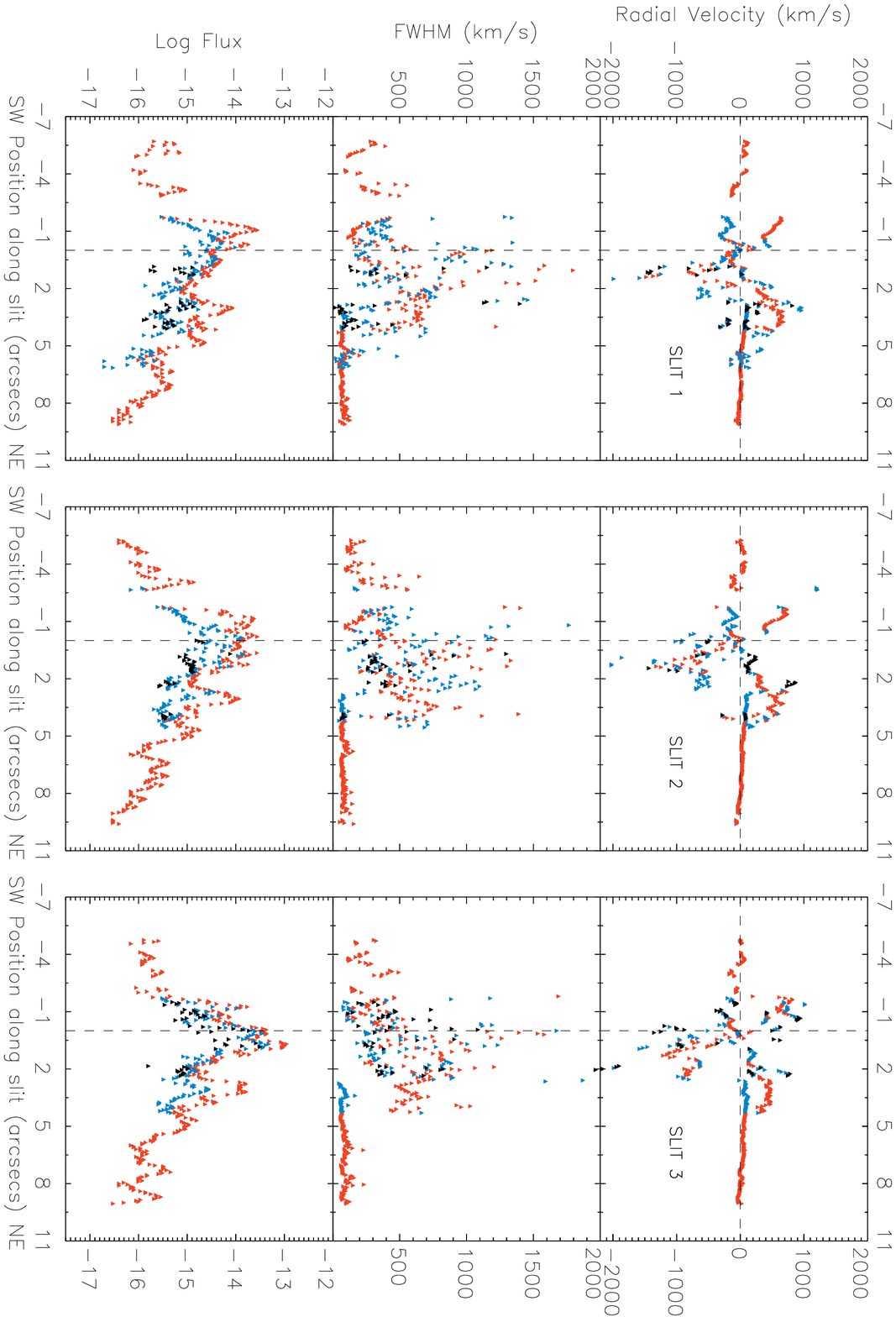]{Plots of radial velocity, FWHM, and flux, from
  slits 1--3, showing multiple components at each position along the
  slit. Red represents the highest flux component, blue the medium
  flux component, and black the lowest flux component for each
  position. An example of the separation is shown in
  \fig\ref{spectra}.\label{vel_fwhm_flux1} }
\figcaption[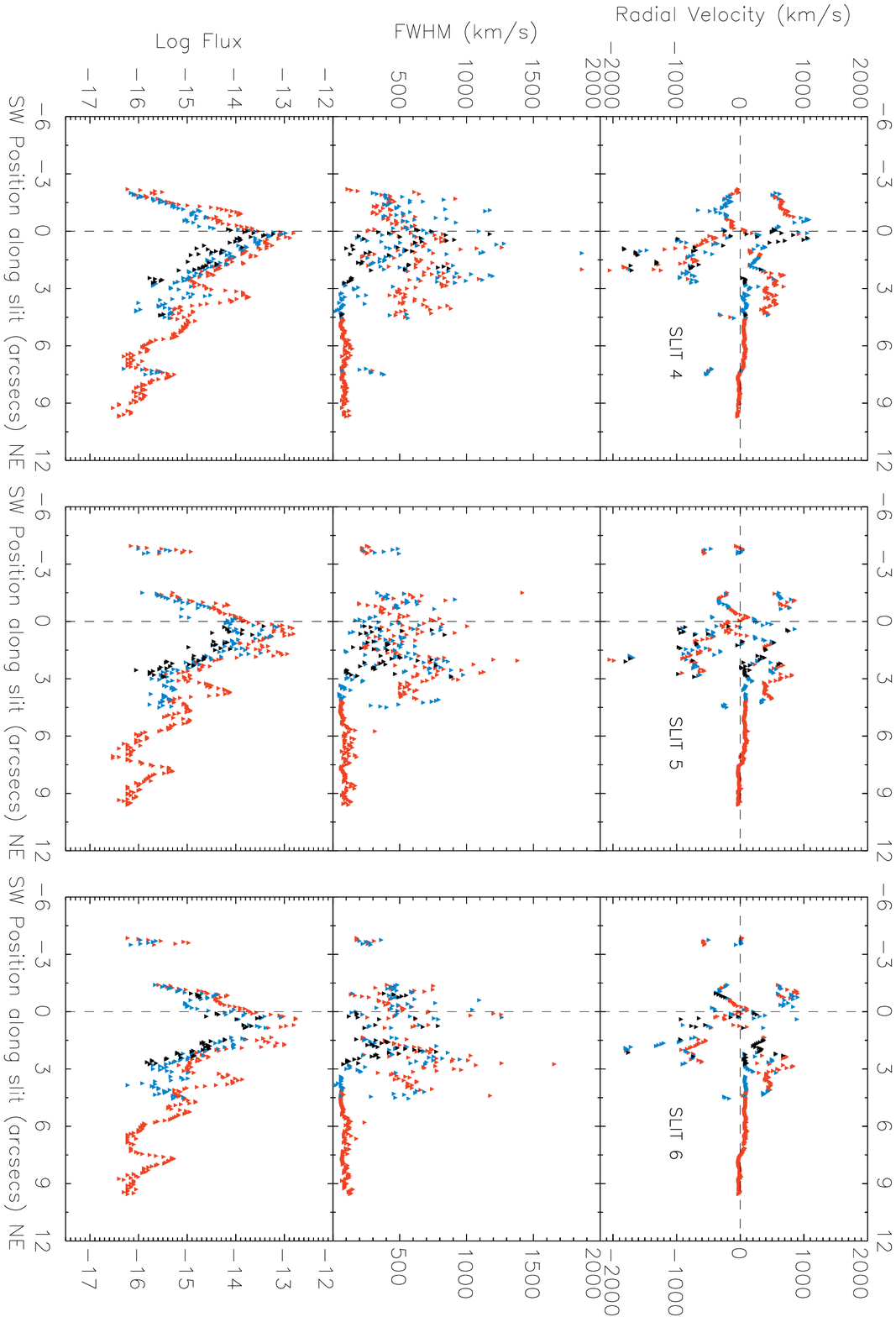]{Same as in \fig\ref{vel_fwhm_flux1}, but for slits
  4--6. \label{vel_fwhm_flux2} }
\figcaption[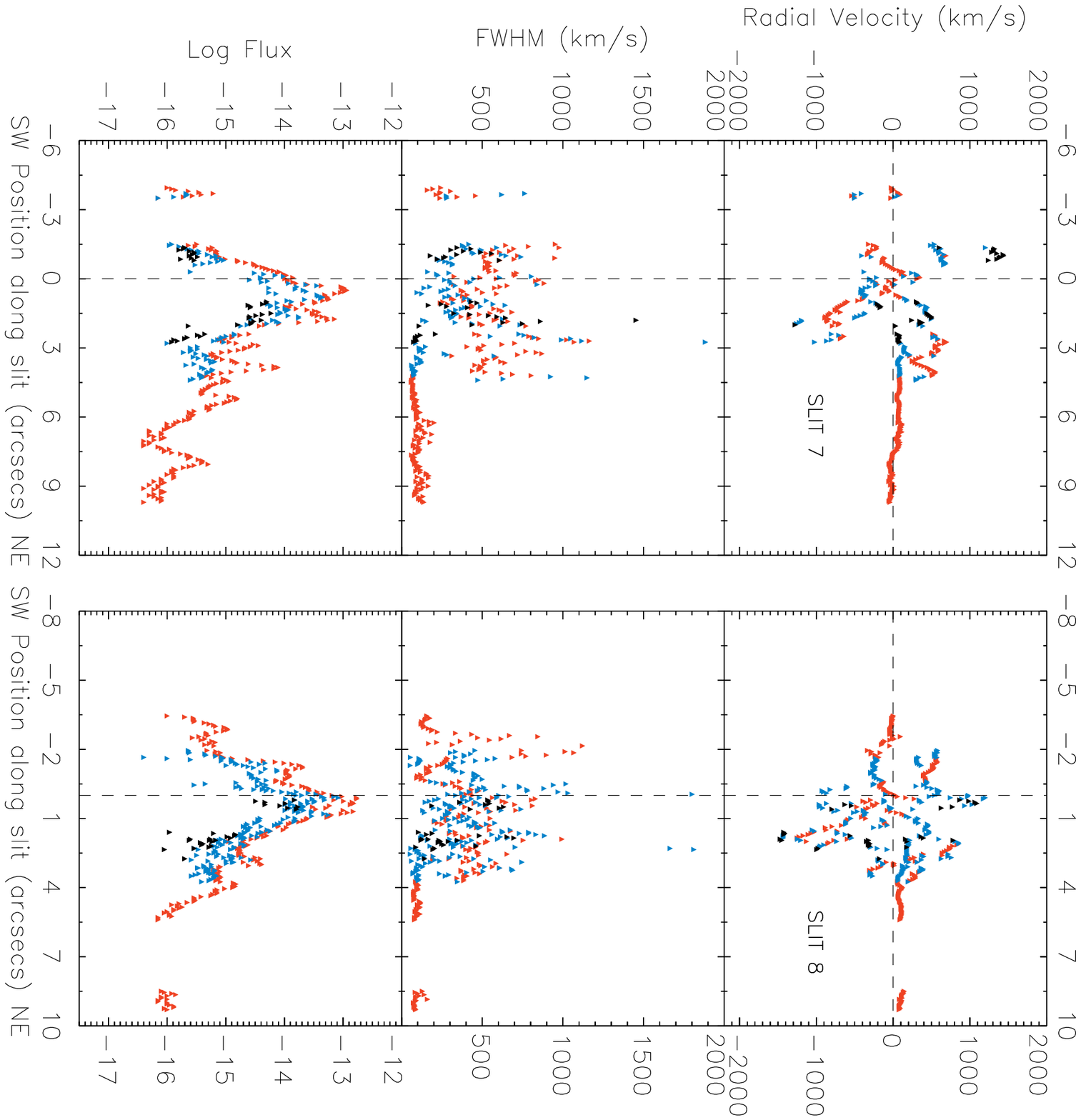]{Same as in \fig\ref{vel_fwhm_flux1}, but for slits
  7--8. \label{vel_fwhm_flux3} }
\figcaption[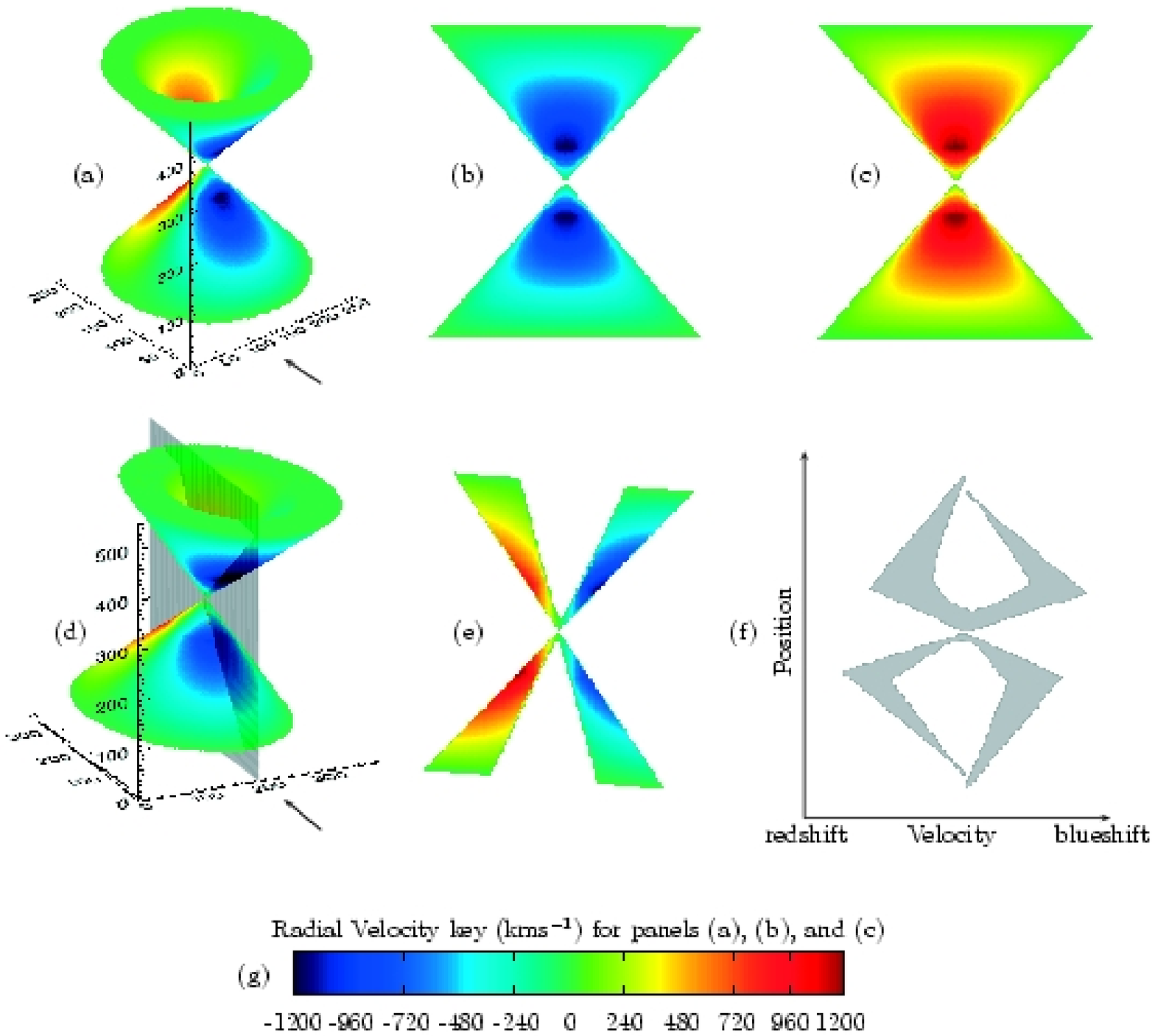]{Bicone models showing the velocity fields for two
  slightly different inclinations. Panel (a) shows the 3-D structure
  of the model with no inclination with the arrow indicating the
  line-of-sight. Panels (b) and (c) show the front and back part
  respectively of the (a) as it would appear in the sky in terms of
  radial velocity measurements. Panels (d), (e), and (f) show our
  kinematic model for \gal\ and our
  slit extraction procedure. The shaded plane in panel (d) mimics the
  position of slit 4 and panel (e) shows the extracted velocities
  along that plane. Panel (f) is a corresponding shaded plot of
  velocities from (e). Panel (g) gives a radial velocity key for models
  (a), (b), and (c). The inclination of (d) changes its maximum
  redshift and blueshift from that indicated by the key; see section \ref{Models} for more
  details.\label{models}}
\figcaption[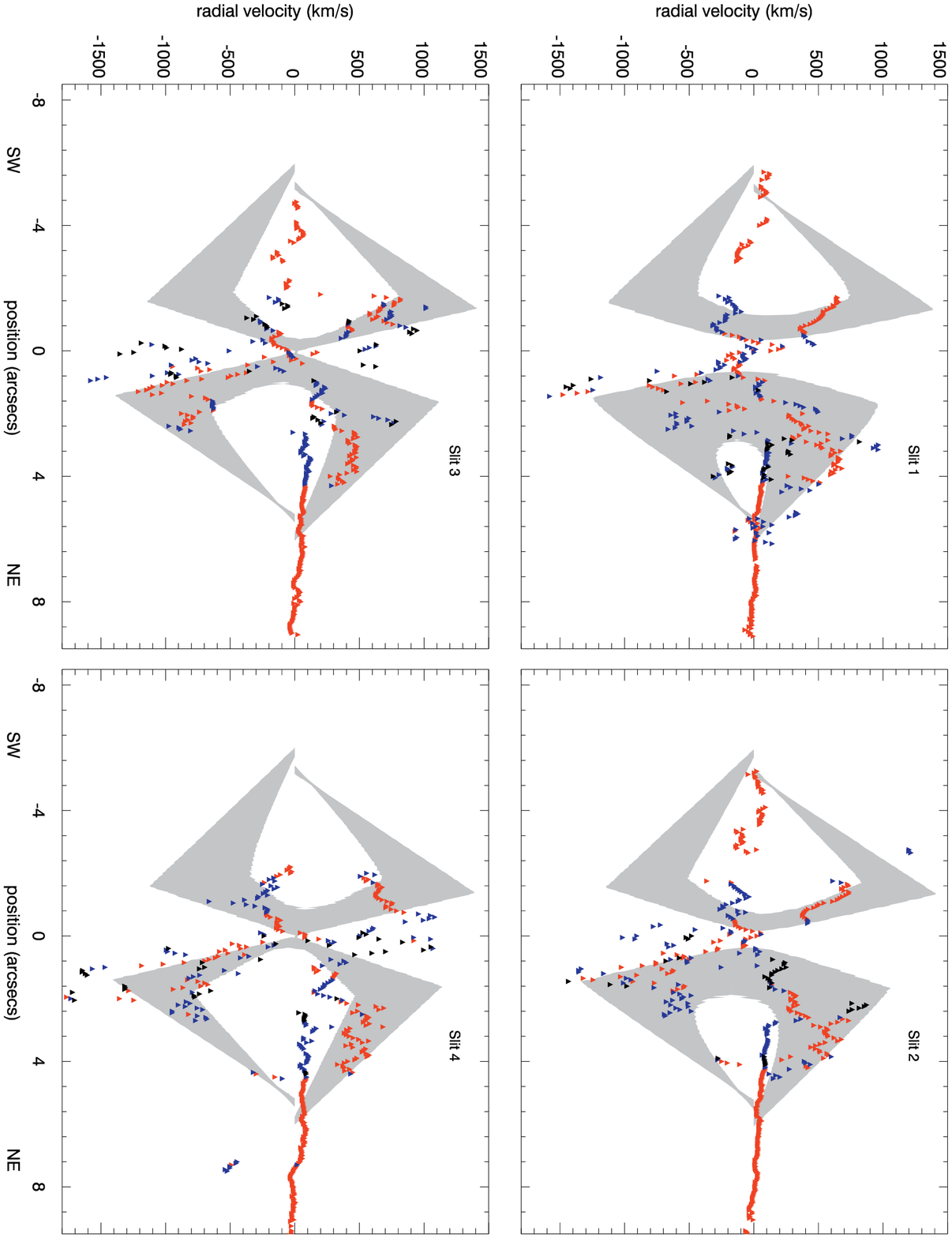]{Models in shaded grey and data points in colors
  showing the fit for slits 1--4. The southwest side of the cone is
  likely extincted by dust in the plane of the host galaxy. The continuous string of
  points in the northeast near 0 \kms, as well as similar points in
  the southwest at many positions, are velocities from
  clouds in the host galaxy. The colors red, blue, and black are
  defined similarly as in Figures~\ref{spectra} and
  \ref{vel_fwhm_flux1}. \label{plotshade1}}
\figcaption[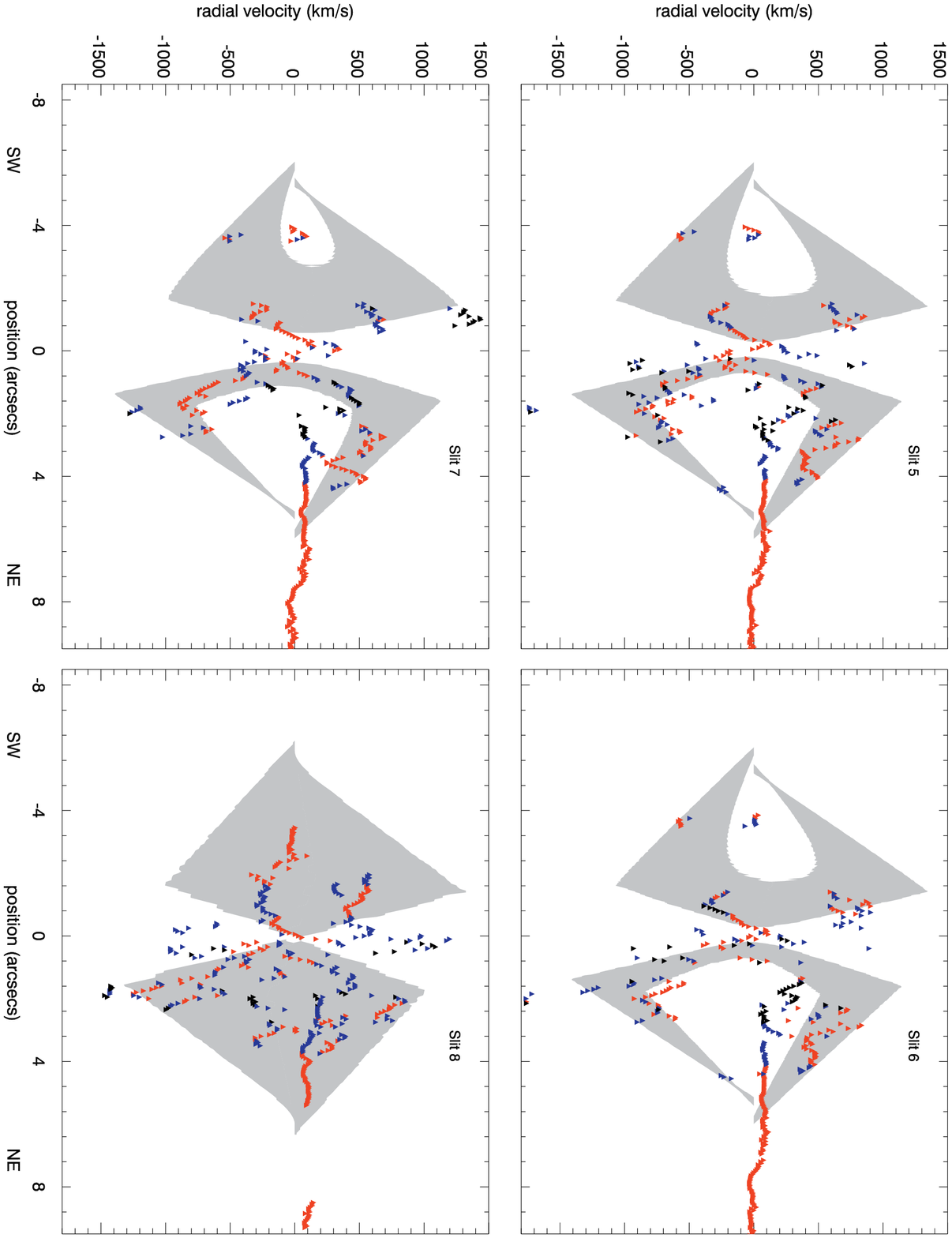]{Same as in \fig\ref{plotshade1} but for slits
  5--8. Slits 5 and 6 differ only by a vertical offset. The small differences
  in some of the radial velocity data points may be due to noise and
  in the way we
  separate the various kinematic components. Slit 8 shows a wide range
  of velocities both by model and data. The low and medium flux points
  at a range of velocities near 0\arcsec\ are outside of our bicone
  geometry, as discussed in the text. \label{plotshade2}}
\figcaption[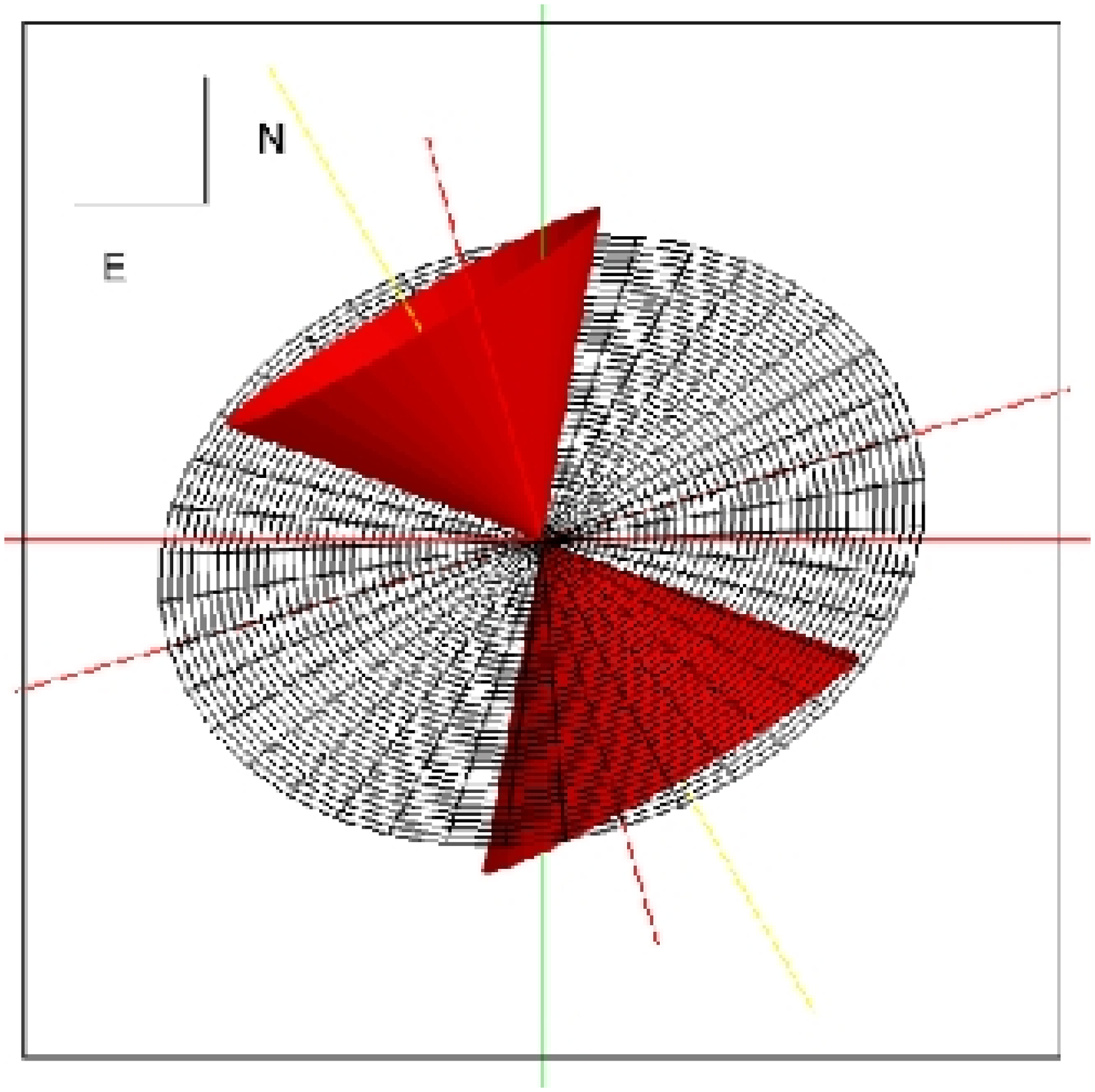]{Plot showing the orientation of the bicone
  relative to the host galaxy as they would appear in the sky. Parts
  of the southwest bicone are being blocked by the galactic disk
  of \gal, hence the lack of data points in the southwest in
  Figures~\ref{plotshade1} and \ref{plotshade2}. \label{host_galaxy}}
\figcaption[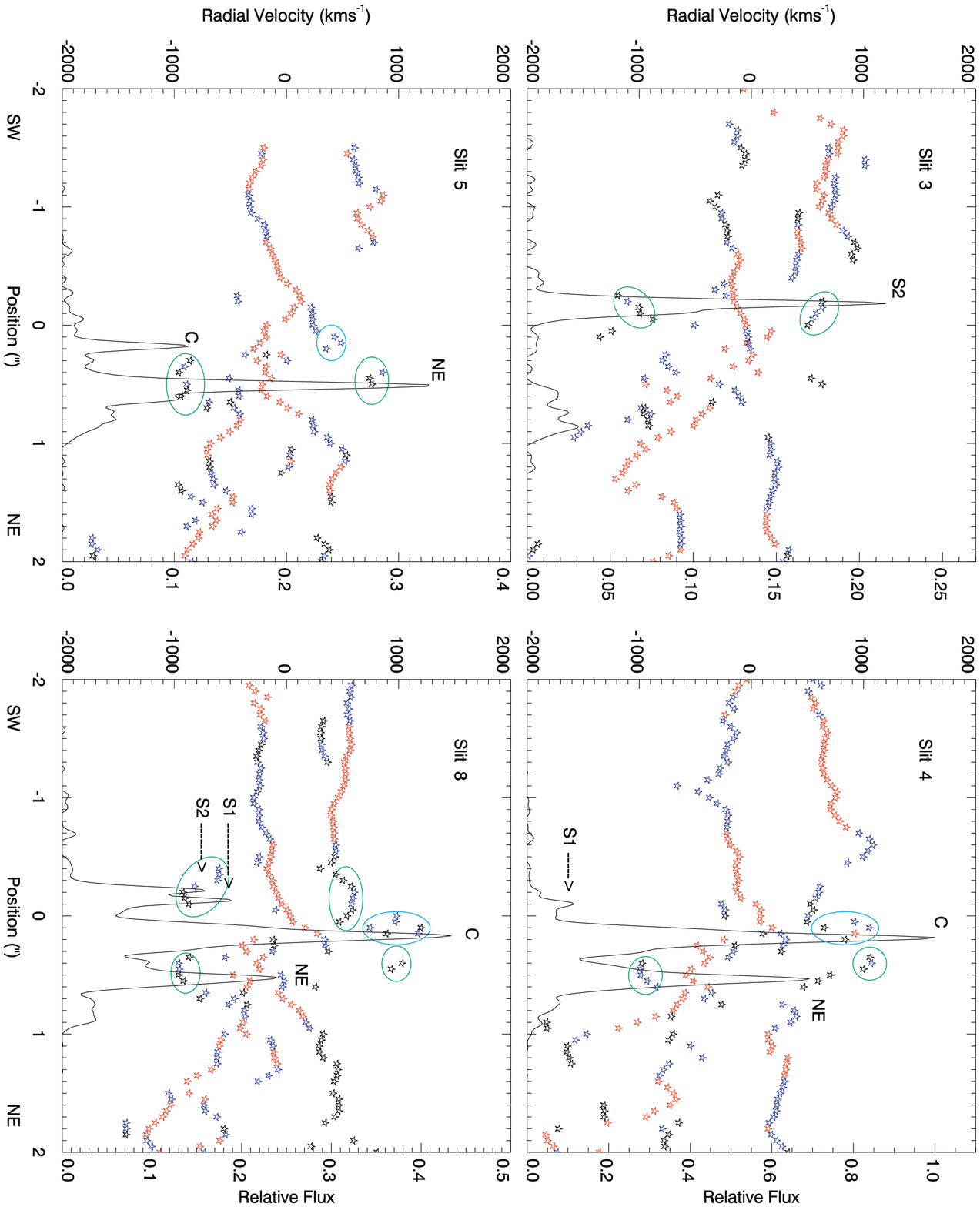]{Plots of relative radio intensities superposed on \OIII\ radial
  velocity plots. The colors of the velocity points are defined as
  in \fig\ref{spectra}. These are the only slits that intercept the
  radio. \label{radio_velocity}}
\figcaption[f12.eps]{Cartoon showing the bicone geometry of the NLR
  with the radio jet in the center in dark color. Lateral expansion
  due to the radio knots is driving fainter NLR clouds perpendicular to the
  bicone axis, independent of the large scale flow of the bright
  clouds. The apex of the bicone therefore becomes broadened.\label{jet_expansion}}

\begin{deluxetable}{lc}
\tablenum{1}
\tablewidth{0pt}
\tablecaption{\textsc{Definition of Model Parameters} \label{model-parameters}}
\tablehead{
\colhead{Parameter} & \colhead{Symbol}}
\startdata
Maximum distance along bicone axis & $ z_{\mathrm{max}} $\\
Inner opening angle                & $ \mathrm{\theta_{inner}} $\\
Outer opening angle                & $ \mathrm{\theta_{outer}} $\\
Inclination of bicone axis         & $ i_{\mathrm{axis}} $\\
Position angle of bicone axis      & $ PA_{\mathrm{axis}} $\\
Maximum velocity\tablenotemark{a}  & $ v_{\mathrm{max}} $\\
Turnover distance\tablenotemark{a} & $ r_{\mathrm{t}} $\\
\enddata
\tablenotetext{a}{deprojected: i.e relative to the nucleus, not the observer}
\end{deluxetable}

\clearpage
\begin{deluxetable}{lcc}
\tablenum{2}
\tablewidth{0pt}
\tablecaption{\textsc{Kinematic Model of NGC 1068} \label{compare-ngc1068}}
\tablehead{
\colhead{Parameter} & \colhead{Previous\tablenotemark{a}} &
\colhead{This Paper}}
\startdata
$ z_{\mathrm{max}} $(pc) & 306 & 400 $ \pm $ 16 \\
$ \mathrm{\theta_{inner}} $(deg)    & 26  & 20  $ \pm  $ 2 \\
$ \mathrm{\theta_{outer}} $(deg)    & 40  & 40  $ \pm  $ 2 \\
$ i_{\mathrm{axis}} $(deg)          & 5\tablenotemark{b} & 5 $ \pm  $ 2\tablenotemark{b}\\
$ PA_{\mathrm{axis}} $(deg)         & 15  & 30  $ \pm  $ 2 \\
$ v_{\mathrm{max}} $(\kms)          & 1300 & 2000 $ \pm  $ 50 \\
$ r_{\mathrm{t}} $(pc)              & 137  & 140  $ \pm  $ 10 \\
\enddata
\tablenotetext{a}{\citet{Crenshaw1068}}
\tablenotetext{b}{northeast is closer}
\end{deluxetable}

\begin{figure}
\epsscale{1}
\vspace{-2cm}
  \plotone{f1.ps}
  \\ Fig. 1
\end{figure}

\clearpage
\begin{figure}
\epsscale{1}
  \plotone{f2.ps}
  \\[-1cm] Fig. 2
\end{figure}

\clearpage
\begin{figure}
  \plotone{f3.ps}
  \\ Fig. 3
\end{figure}

\clearpage
\begin{figure}
  \plotone{f4.ps}
  \\ Fig. 4
\end{figure}

\clearpage
\begin{figure}
  \plotone{f5.ps}
  \\ Fig. 5
\end{figure}

\clearpage
\begin{figure}
  \plotone{f6.ps}
  \\ Fig. 6
\end{figure}

\clearpage
\begin{figure}
\epsscale{1}
  \plotone{f7.ps}
  \\[-1cm] Fig. 7
\end{figure}

\clearpage
\begin{figure}
  \plotone{f8.ps}
  \\ Fig. 8
\end{figure}

\clearpage
\begin{figure}
  \plotone{f9.ps}
  \\ Fig. 9
\end{figure}

\clearpage
\begin{figure}
%\epsscale{1}
  \plotone{f10.ps}
  \\ Fig. 10
\end{figure}

\clearpage
\begin{figure}
%\epsscale{1}
  \plotone{f11.ps}
  \\ Fig. 11
\end{figure}

\clearpage
\begin{figure}
%\epsscale{1}
  \plotone{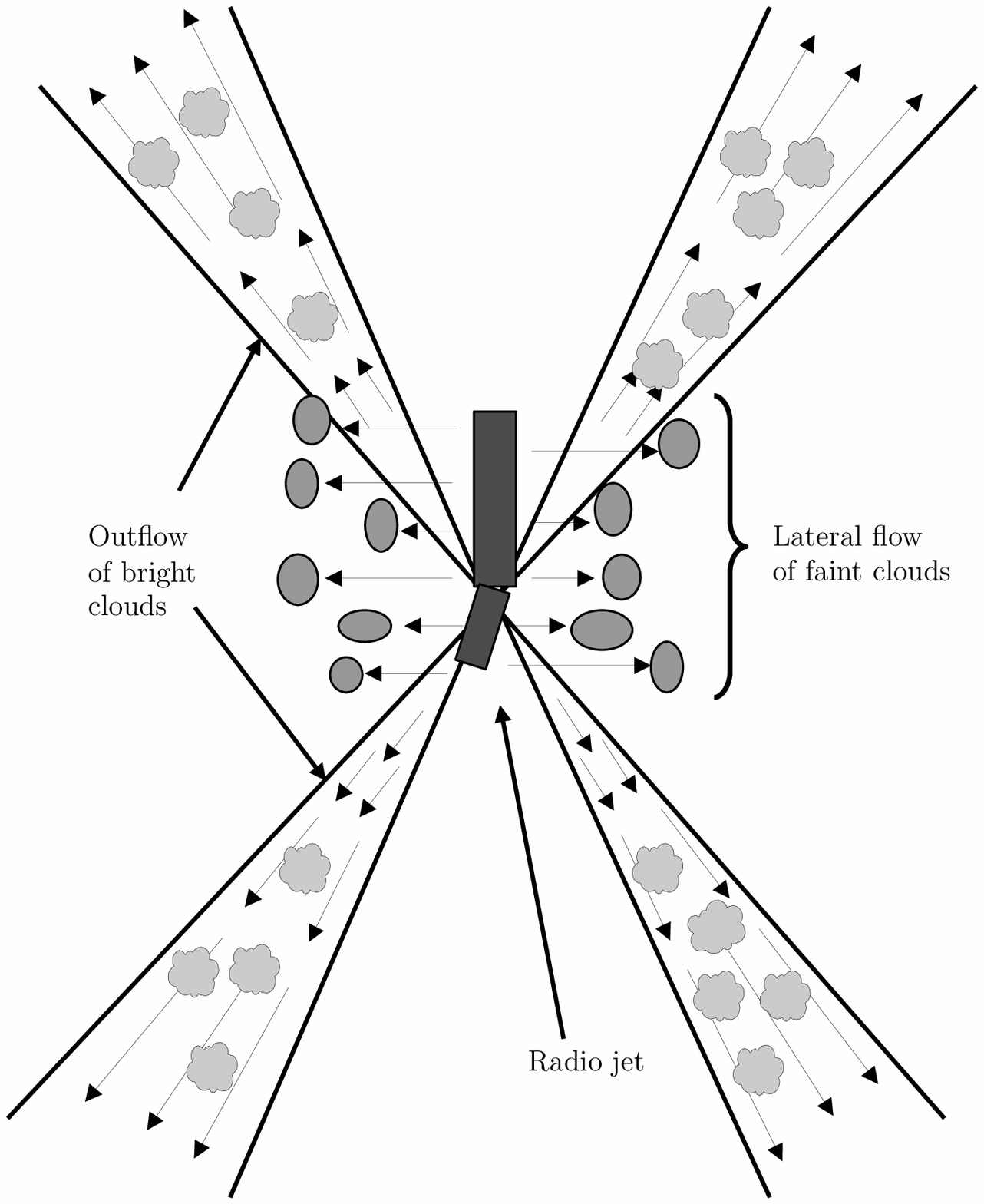}
  \\ Fig. 12
\end{figure}

\end{document}